\documentclass[runningheads]{llncs}
\usepackage{graphicx}
\usepackage{xcolor}
\usepackage[english]{babel}
\usepackage{csquotes}
\usepackage{cite}
\usepackage[caption=false]{subfig}
\usepackage{float}
\PassOptionsToPackage{hyphens}{url}
\usepackage{hyperref}
\usepackage{enumitem}

\begin{document}
\title{A Short Survey on Business Models of Decentralized Finance (DeFi) Protocols}

\titlerunning{A Short Survey on Business Models of DeFi Protocols}

\author{Teng Andrea Xu\inst{1} \and
Jiahua Xu\inst{2, 3}}
\authorrunning{T. Xu et al.}
%
\institute{École Polytechnique Fédérale de Lausanne, Lausanne, Switzerland, \email{andrea.xu@epfl.ch} \and
University College London, London, UK, \and The DLT Science Foundation, London, UK, \\
\email{jiahua.xu@ucl.ac.uk}\\
}
\maketitle              
\begin{abstract}
Decentralized Finance (DeFi) services are moving traditional financial operations to the Internet of Value (IOV) by exploiting smart contracts, distributed ledgers, and transactions among different protocols. The exponential increase of the Total Value Locked (TVL) in DeFi foreshadows a bright future for automated money transfers in a plethora of services. In this short survey paper, we describe the business models of various DeFi protocol types---namely, \textit{Protocols for Loanable Funds (PLFs)}, \textit{Decentralized Exchanges (DEXs)}, and \textit{Yield Aggregators}. We then abstract the general business models of those protocol types and compare them. Finally, we provide open research challenges that will involve different domains such as economics, finance, and computer science. 

\end{abstract}

\begin{figure}[tb]
    \centering
    \subfloat[ \label{fig:uni}]{
        \includegraphics[width=0.3\linewidth]{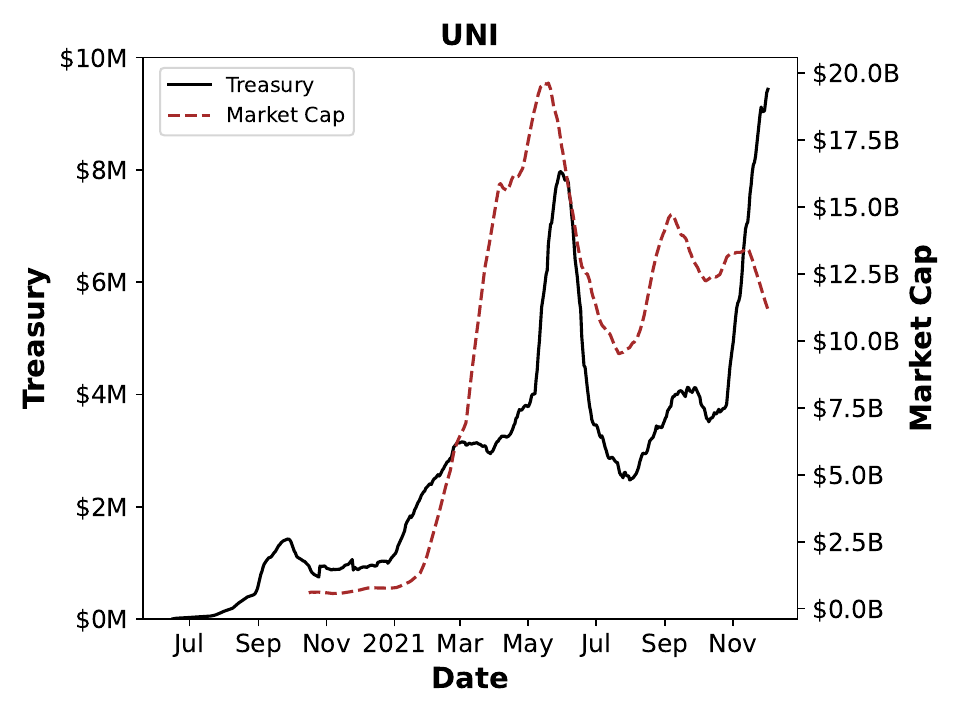}
    }
    \subfloat[ \label{fig:yearn}]{
        \includegraphics[width=0.3\linewidth]{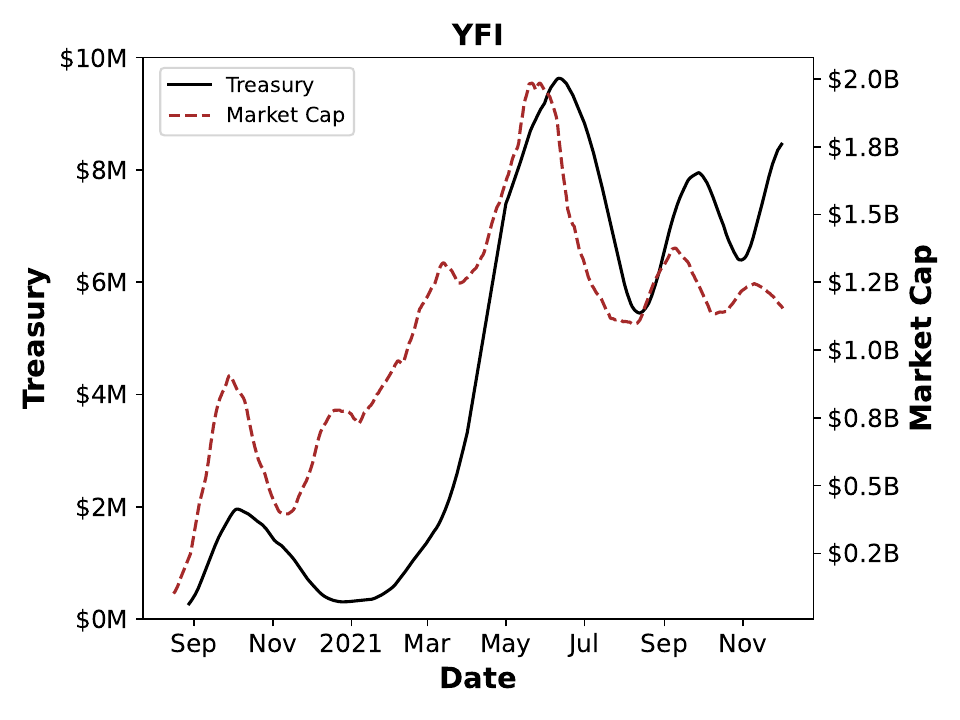}
    }
    \subfloat[\label{fig:aave}]{
        \includegraphics[width=0.3\linewidth]{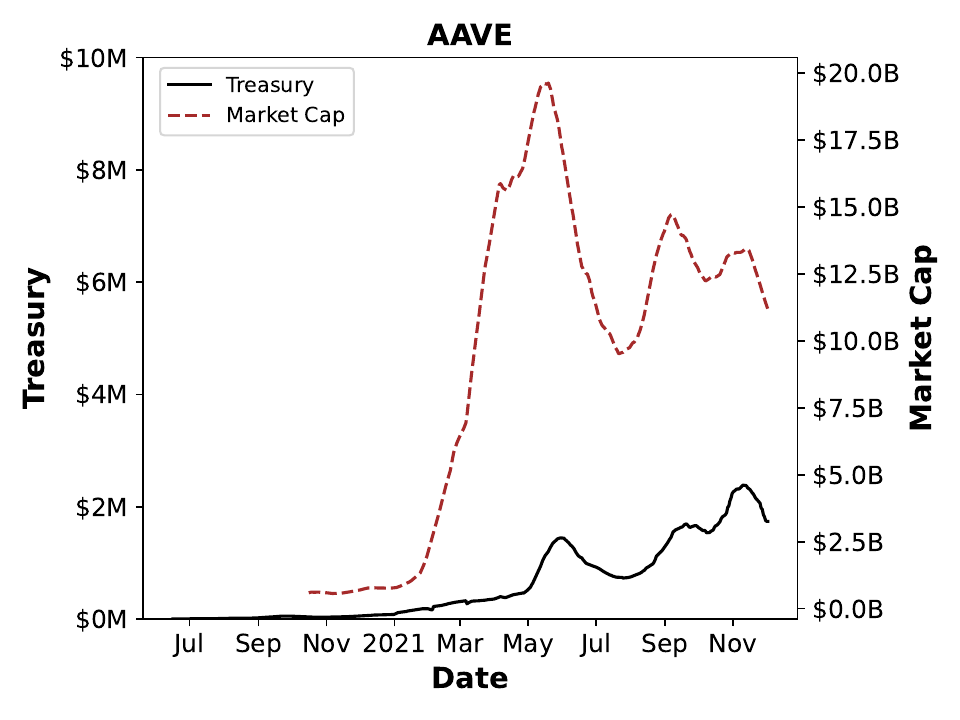}
    }
    \caption{The figure above shows the top DeFi's protocols daily treasury size, in solid black, and market cap, in dashed red, for each financial service. We select each token according to DeFiPulse TVL. Both measures are smoothed by using a rolling window of 30 days. To date, Uniswap---by default---is not applying any exchange fee but its treasury consists of 43\% of the total token supply \cite{uniswap_treasury}, whereas AAVE and Yearn apply fees for their financial services \cite{aave_fee, yearn_fee}. We can see that market cap drives financial services usage, and thus protocols' revenues. Uniswap and AAVE data are retrieved from CryptoFees API \cite{cryptofees}, while YFI data is retrieved from YFIstats \cite{yfistats}. The respective correlations between market cap and treasury size are: (a) 0.596, (b) 0.857, and (c) 0.431.} 
    \label{fig:revenue}
\end{figure}
\section{Introduction}

Decentralized Finance (DeFi) aims to provide financial services on a blockchain-based infrastructure. A plethora of protocols form the DeFi ecosystem. These protocols are able to replicate classical financial services, such as lending and exchange, without any central institution, via smart contracts and the immutable ledger. In the literature \cite{werner2021sok,schar2021decentralized}, DeFi's key features are generally recognized to be \textit{open} to anyone, \textit{transparent}, \textit{non-custodial}, and \textit{composable}, i.e. financial services can be arbitrarily composed to make new financial products. The Total Value Locked (TVL)\footnote{ The TVL is commonly defined as the sum of all assets' value, denominated in USD, deposited in a DeFi protocol, and therefore locked in the underlying set of smart contracts.} has seen exponential growth with the so-called \enquote{DeFi Summer}. TVL grew from \$600m as at the end of March 2020 to \$11bn as at the end of September 2020 \cite{cousaert2021sokyearn}. Furthermore, from March 2020 to the time of writing\footnote{2021-10-31, https://defipulse.com/}, the top-100 DeFi tokens' market cap grew by almost a hundred times---from \$1.8bn to \$154bn \cite{coingecko_market_cap}. These recent DeFi milestones foreshadow a bright future for both DeFi's users and investors. We show, in Figure \ref{fig:revenue}, the top Ethereum DeFi tokens' daily treasury size together with their market cap, smoothed by a rolling window of 30 days. We show a plot for each DeFi financial service, i.e. Uniswap (UNI) for \textit{DEXs}, Yearn Finance (YFI) for \textit{Yield Aggregators}, and AAVE for \textit{PLFs}. 
To date, there is a lack of literature that offers any clear abstraction on how DeFi protocols generate their revenue stream, a key component for the sustainability of a project. We claim that it is important for both investors and users to understand how DeFi tokens profit. From an investor's perspective, a clear business model with a steady and constant revenue stream are key features before investing in the underlying project. As an end user, DeFi's users look for reliable protocols; hence, a protocol with an efficient and observable business model is likely to be a \enquote{secure} protocol. Therefore, the central contribution of this short survey paper is to describe and offer a clean comparison of different DeFi services business models.  In this work, we will look into the main DeFi financial services namely: \textit{Protocols for Loanable Funds (PLFs)}, \textit{Decentralized Exchanges} (DEXs), and \textit{Yield Aggregators}. We refrain from analyzing the variety of protocols in technical detail, but rather direct the reader to other resources. The focus of this work is mainly on the protocols’ business model. The paper is structured as follows. First, we describe the general PLFs' business model in Section \ref{sec:plfs}. Subsequently, we explain the dominant cash flows within DEXs and Yield Aggregators in Section \ref{sec:dex} and \ref{sec:yield}. In Section \ref{sec:defi}, we present a first generalized business model in DeFi. Finally, we provide a literature review and conclude the work with Section \ref{sec:literature} and \ref{sec:conclusion}.

\section{Major DeFi Protocols}

\subsection{Protocols for Loanable Funds (PLF)}
\label{sec:plfs}

\begin{figure}[tb]
    \centering
    \includegraphics[width=0.6\linewidth]{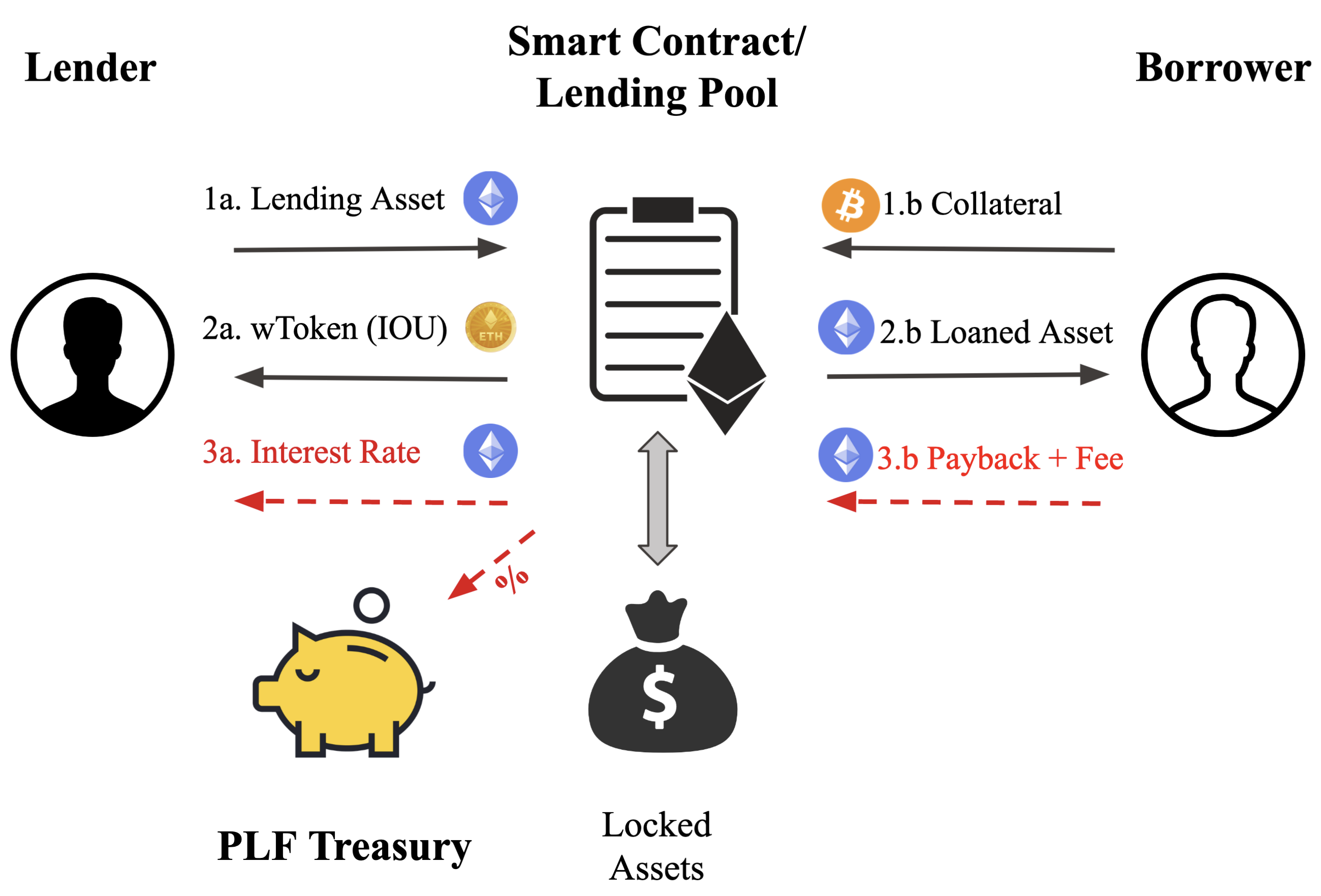}
    \caption{The figure above abstracts and generalizes the lending protocol framework by showing the main actors and interactions. From left to right, lenders can deposit their crypto-assets to gain additional profits. They receive a PLFs wrapped token or IOU as proof of their deposit. In the center, the smart contract takes care of the deposited assets, loans, and liquidations---if any. On the right, a borrower must deposit collateral before getting the loan. Finally, at the end of the loan, the borrower will have to return the borrowed amount plus an interest; part of this interest will be redistributed pro-rata to all lenders, and the rest will generate revenue for the PLF itself.}
    \label{fig:lending}
\end{figure}

PLFs let users borrow/lend digital assets in a decentralized fashion. Automated smart contracts behave as middle-men. They lock assets deposited by the lenders and allow borrowers to get liquidity in exchange for collateral \cite{perez2021}. These types of smart contracts are also called Lending Pools \cite{qin2021}. These Lending Pools typically lock a pair of tokens, a loanable token, and a collateral token. By providing liquidity, lenders gain interest revenue depending on the supply and demand. Because there is no guarantee of repayment, borrowers must over-collateralize their position \cite{vadgama2022}. On top of that, when returning the amount borrowed, the borrowers must pay interest expenses that are redistributed proportionately to the lenders and the governance token holders. When a borrower's loan position becomes liquidated, they will have to pay an additional fee. We show in Figure \ref{fig:lending} a typical generalized PLF use case.

\subsubsection{Business Model}
PLFs' cash flow depends on the interest rate model, the current underlying demand versus supply, and the total amount borrowed. The interest rate model can be either a linear model, a non-linear model, or a kinked model \cite{gudgeon2020plfs}. The interest rate is driven by the underlying asset demand-supply ratio: the interest rate is higher when the ratio is high, and vice versa. Given the rate of interest, the PLF gets a percentage of it. For example, Compound takes 10\% of the interest rate \cite{compound_interest}. Besides traditional over-collateralized loans, PLFs offer flash loans\footnote{Flash Loans are a special type of loan where the borrower must return the borrowed amount plus interest in the same transaction without the need for collateral. An in-depth explanation can be found in \cite{qin2021attacking}.} that can bring more revenues to the protocol. Flash loans interest fee is usually fixed, e.g. AAVE \cite{aave_flash_loan}, or even without fees, e.g. dYdX \cite{dydx_flash_loan}.

\begin{figure}[tb]
    \centering
    \includegraphics[width=0.6\linewidth]{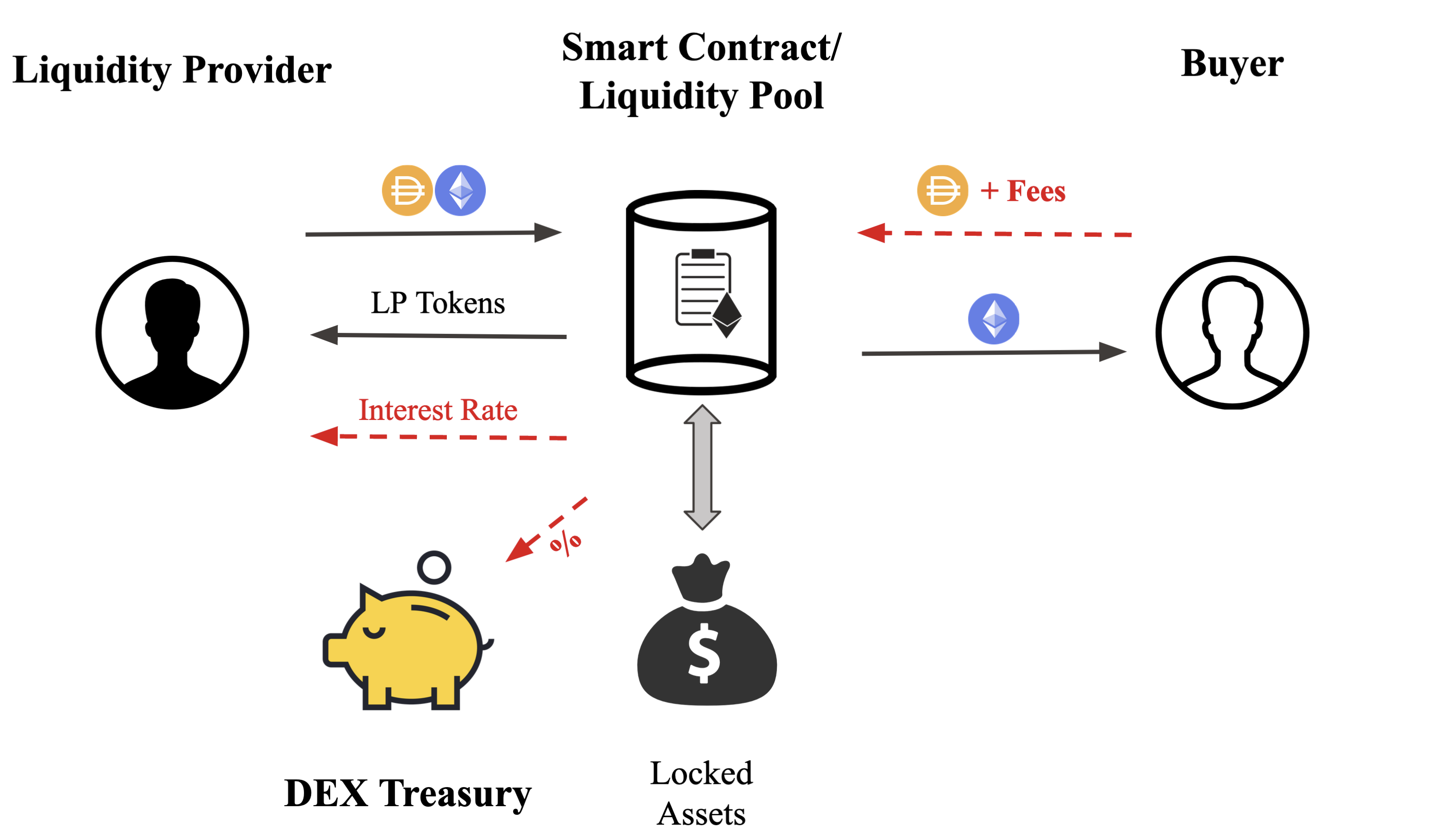}
    \caption{The figure above abstracts and generalizes DEXs protocols framework by showing the main actors' interactions and the protocol revenue stream. From left to right, LPs deposit a pair of tokens, in this example DAI/ETH, in the Liquidity Pool. In exchange, they receive LP tokens as proof of their deposit. In the middle, a smart contract takes care of locked assets, new deposits, swaps, and fees. The buyer will have to pay a fee for his swap. This fee will be partially distributed pro-rata among all LPs, while the DEX treasury will collect a percentage of it.}
    \label{fig:dex}
\end{figure}

\subsection{Decentralized Exchanges}
\label{sec:dex}
DEXs operate differently to  classical order-book exchanges where traders match market bids and/or asks. Again, smart contracts are the middle-men and, in this case, are called Liquidity Pools. Investors or, in this scenario, Liquidity Providers (LPs) can deposit, in the case of constant product pools \cite{vitalik_invariant}, a pair of equal worth tokens---say ETH/DAI as shown in Figure \ref{fig:dex}---into these Liquidity Pools. In exchange, they will receive LP tokens as proof of their deposit and earn a percentage of the fee accrued from the buyer when swapping. A price is assigned for each token given the protocol's price function. The buyer that is willing to swap DAI for some ETH will deposit DAI in the Liquidity Pool, plus some fee, and receive ETH. The whole mechanism is called Automatic Market Making (AMM). Further reading on the topic can be found in \cite{xu2021sokdex, lin2019deconstructing}. We show in Figure \ref{fig:dex} a typical generalized DEX use case.

\subsubsection{Business Model}
When buyers swap, they pay a fee. This fee is split pro-rata between the liquidity providers of the pool as a reward for their contribution to the pool. A percentage of the fee is sent to the protocol's treasury. This share of the fee represents the primary income resource for most AMMs. Balancer, for example, has recently increased its protocol fee\footnote{Note that the protocol fee is applied on the swap fee.} from 10\% to 50\% \cite{balancer_fee} equaling Bancor's fees \cite{bancor_medium_fee}. To date, even though it doesn't apply any protocol fee by default \cite{uniswap_whitepaper}, Uniswap has the biggest treasury in DeFi with \$9bn locked in its treasury \cite{treasury}, which is the value of the 43\% of Uniswap total supply \cite{uniswap_treasury} locked for the community.

\begin{figure}[tb]
    \centering
    \includegraphics[width=0.6\linewidth]{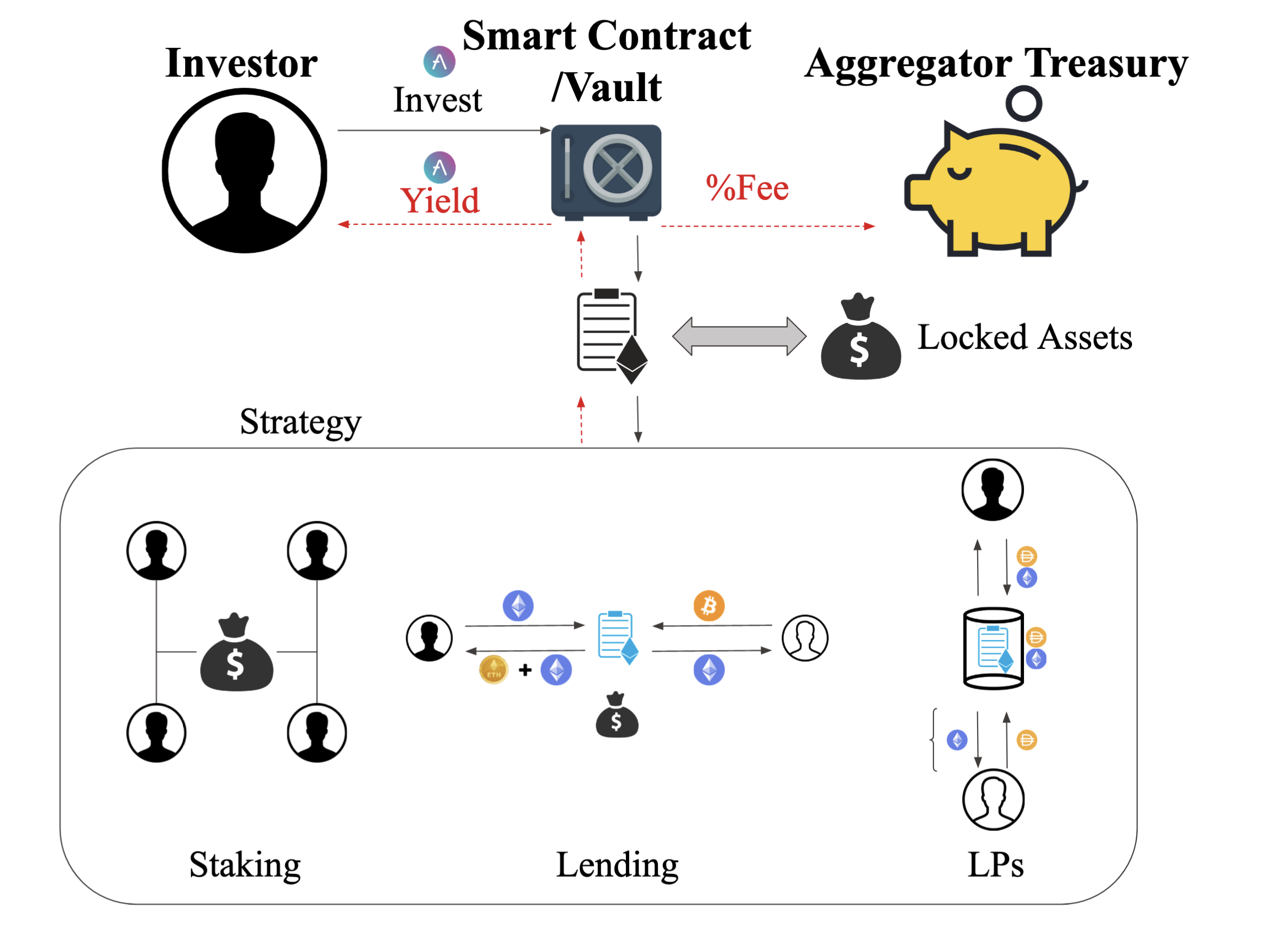}
    \caption{This figure shows a typical yield aggregator use case. Starting from the top left, the investor chooses his Vault of preference to deposit his savings. The Vault, a smart contract, will run its pre-set strategy; simple staking, lending, or providing liquidity are just examples. More complex strategies combine borrowing and/or leveraging involving multiple steps and protocols---yETH vault is an example of a multiple-step strategy \cite{yearn_yeth}. Usually, Vaults apply a fixed performance fee on the strategy yield.}
    \label{fig:yearn_agg}
\end{figure}

\subsection{Yield Aggregators}
\label{sec:yield}
Yield Aggregators combine different strategies to maximize investors' rate of return. Similar to PLFs and DEXs protocols, smart contracts have a central role. Commonly, smart contracts are referred to as \enquote{Vaults} in this domain. In this scenario, shown in Figure \ref{fig:yearn_agg}, investors deposit their savings into a Vault. Different Vaults run different strategies. These strategies can be straightforward, such as finding the best lending protocol interest rate, or more complex, as borrowing assets and leveraging some other position by exploiting different protocols. For a more in-depth technical insight refer to \cite{cousaert2021sokyearn, xu2023reap}.

\subsubsection{Business Model}
Yield Aggregators cash flow is based on their Vaults' performance. That is, yield aggregators charge a commission on the strategy's profit. Hence, the investor yield will be equivalent to the Vault's total profit minus the protocol's fee. Different tokens apply different fee rates: Yearn Finance v2 applies 20\% as performance fee and an additional 2\% as management fee \cite{yearn_fee}, Pickle Finance and Idle have 20\% and 10\% performance fees respectively \cite{pickle_fee, idle_fee}. Harvest is the only yield aggregator that applies 30\% of fees but uses the whole reward to buy back FARM---Harvest native tokens---from DEXs and redistribute them to stakers \cite{harvest_fee}. 



\section{DeFi Business Model}
\label{sec:defi}
\begin{figure}[tb]
    \centering
    \includegraphics[width=0.6\linewidth]{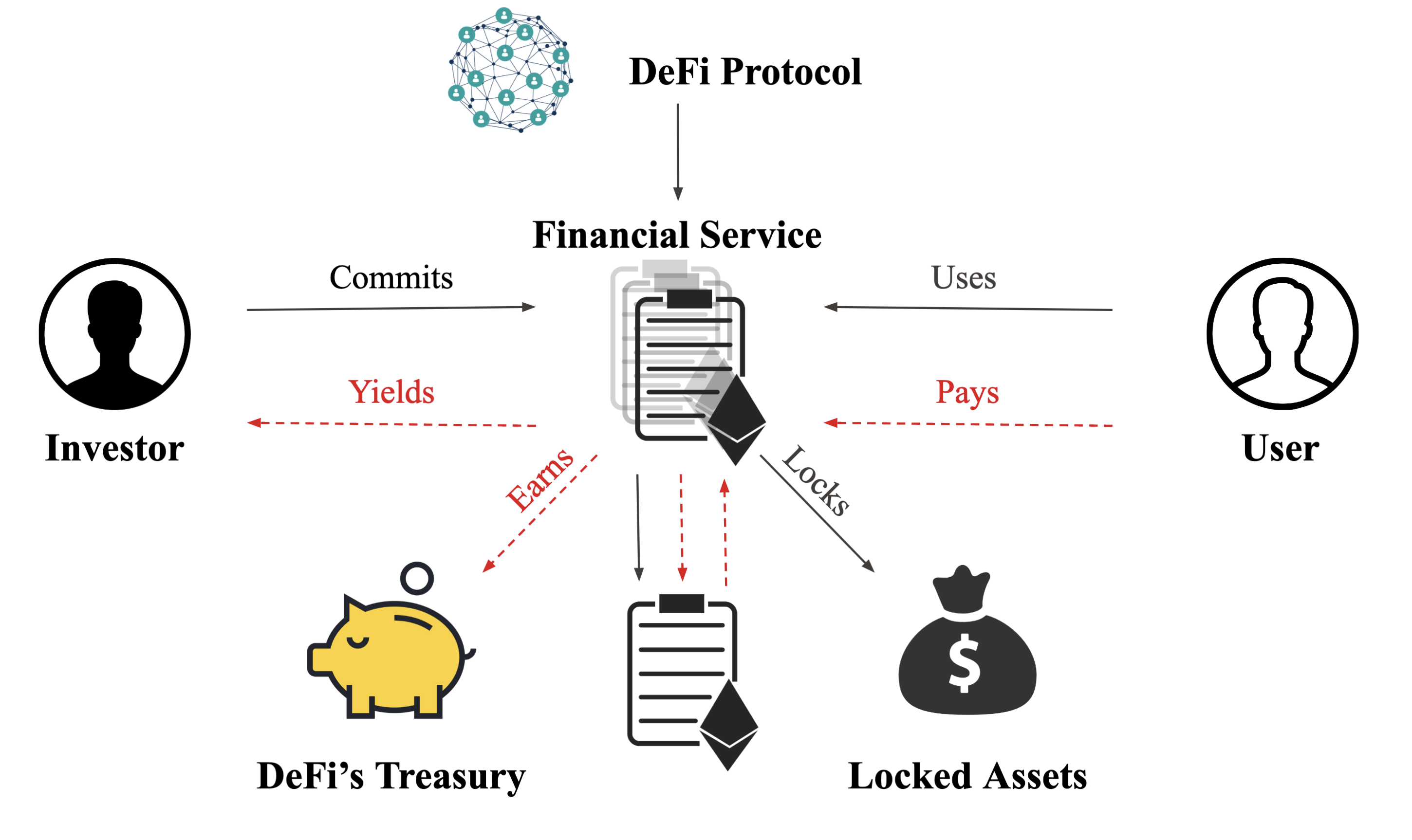}
    \caption{DeFi common mechanism and revenue strategy.}
    \label{fig:generalization}
\end{figure}

We synthesize the business models reported and give a first general DeFi business model framework (Fig. \ref{fig:generalization}). This framework involves different actors and actions with their naming conventions (Table \ref{tab:sok}).

\begin{enumerate}[leftmargin=0pt]
    \item[\bf DeFi Protocol] A set of smart contracts, with multiple facets---PLFs, AMMs, or Yield Aggregators. They provide open, non-custodial, permissionless, and composable financial services in exchange for a small fee. The fee is applied to any asset movement, for example, borrowing assets or swapping assets. 
    \item[\bf Investor] This actor is willing to hold the underlying protocol risk, such as protocol misbehavior, impermanent loss, or rug-pulls, in exchange for a passive income. Therefore, he mainly deposits his assets and provides liquidity to the financial service.
    \item[\bf User] The user usually exploits the protocol on the fly, and he never waits for any long-term response---save for the case of Yield Aggregators, where the user is also an investor. This actor requests asset movements, and thus pays interest rates to the protocol.
    \item[\bf Financial Service] This actor is the core of the whole protocol: Locks assets, satisfies the assets movements requests, and prevents protocol misuse. Furthermore, it can behave as an investor by leveraging other DeFi Protocols. Finally, it delivers yields and earnings to the other actors. 
\end{enumerate}

DeFi's services do not come for free. As we have seen, investors can exploit these services to earn passive income. However, this additional interest is somehow \enquote{taxed} by the DeFi protocol. On the other side, the users are willing to use the platform in exchange for a fee. Hence, the DeFi protocol has an income from both sides. In classical finance, this market and business model is known as the \enquote{Two-Sided Markets}, first formalized in \cite{rochet2003platform}. On one side, the investor provides liquidity to the financial service that peer users can use. On the other side, by paying fees, the user provides income to both protocols and investors.  

\begin{table}[tb]
\caption{An overview of DeFi's naming taxonomy.}
\begin{tabular}{|l|l|ll|l|}
\hline
\textbf{DeFi Protocol}     & \textbf{Smart Contract} & \multicolumn{1}{l|}{\textbf{Investor}}  & \textbf{User} & \textbf{Financial Service} \\ \hline
\textit{PLFs}              & Lending Pool            & \multicolumn{1}{l|}{Lender}             & Borrower      & Loan                       \\ \hline
\textit{DEXs}              & Liquidity Pool          & \multicolumn{1}{l|}{Liquidity Provider} & Buyer/Trader  & Exchange                   \\ \hline
\textit{Yield Aggregators} & Vault                   & Vault User                              &               & Asset Management           \\ \hline
\end{tabular}
\label{tab:sok}
\end{table}

\section{Literature Review}
\label{sec:literature}
Our work is strongly related to the canonical empirical asset pricing problem. Evaluating and pricing an asset is perhaps the most renowned problem in classical finance literature. There exist two cardinal analyses to value and price an asset. On the one hand, technical analysis deduces future underlying value from its history of trading, which can involve price changes, trading volume, price moving average, and other historical characteristics \cite{edawards2018technical}. \cite{taylor1992use} shows how 90\% of chief foreign exchange dealers based in London in November 1988 exploited technical analysis for their trades. We observe here a duality between public equity markets and cryptocurrencies trading. On the other hand, fundamental analysis is based on a firm's book value---assets and liabilities--- stream of dividends, current earnings, future investment opportunities, etc. \cite{modigliani1961dividend} studies the role of dividends policies in a firm's share evaluation. The paper shows how firm's value can be explained by its current earnings, the growth rate of earnings, the internal rate of return, and the market rate of return. Remarkable findings in the series of work \cite{Fama1993, fama1992cross, fama1995size}, show how firm book value and market equity size are important to explain future company's return. Fundamental analysis is also known as \enquote{value investing} \cite{piotroski2000value}, or \enquote{intrinsic valuation} \cite{damodaran2012}. Similarly, do high-value tokens perform better than low-value cryptocurrencies as high-value stocks did in the past? Finally, we would like to stress the parallel among firms' share ownership and DAO tokens \cite{xu2023defi}. Both of them give shareholders voting power on underlying future actions. We will address and expand on these open research challenges in the next Section \ref{sec:conclusion}.

\section{Conclusion}

In this short survey paper, we synthesize DeFi's main business models. We claim that it is important for DeFi's investors and users to understand which protocol has a reliable cash flow. Moreover, the literature is missing a clear overview of the protocols' business model to date. The main contributions of this short survey paper are as follows. First, we provide a clear understanding and explanation of the most important DeFi's services business model. Furthermore, we establish a novel general framework business model adopted by DeFi protocols. On top of that, the scientific community can address multiple open research challenges:

\begin{enumerate}[leftmargin=0pt]
    \item[\bf Value Investing] In classical finance, a firm has a high \textit{value} when it has a high book-to-market ratio. In the literature, it is shown that value stock portfolios achieve a higher long-term mean return \cite{piotroski2000value}. Conducting a parallel study on crypto-assets is a straightforward application of Value Investing.
    \item[\bf Voting Power] DAO tokens grant the holder voting power on the underlying protocol changes, similar to public company shareholders. However, these tokens don't have an initial value, but they acquire it by exchanging and trading. The price discovery of similar DAO tokens is also an open research question.
    \item[\bf Regulatory Issues] Recently, the cryptocurrency ecosystem has seen significant turbulence. While, China has completely banned all crypto transaction \cite{bbc_china_ban}, India is working on plans to enforce similar regulations \cite{bbc_india_ban}. Finally, the US is willing to regulate the market \cite{time_us_regulation}. We claim that the literature lacks a deep analysis that evaluates DeFi's business models suitability with the current regulatory framework.
\end{enumerate}

DeFi's financial services have seen massive growth since the \enquote{DeFi Summer}. While some protocols achieved a stable cash flow, many others have been subject to cyber-security breaches \cite{cousaert2022}. In 2021 only, the whole DeFi world has suffered almost \$1bn loss due to hacks \cite{cointelegraph_hacks}. The most recent hack is dated December 1st, 2021. BadgerDAO, a service that allows Bitcoin to be used as collateral over different DeFi protocols, suffered a \$120m loss \cite{coindesk_badger}. Albeit the exponential growth in DeFi's users and revenue stream, DeFi is yet a risky infrastructure that has to mature over time. Whether DeFi will replace or co-exist with the classical financial services is yet unclear, leaving us with open research challenges to unveil.

\label{sec:conclusion}
%
%
%
%

\end{document}